\documentclass[sigconf]{acmart}
\usepackage[utf8]{inputenc}			
\usepackage[english]{babel}			
\usepackage{duran_esem_rr_2021}	

\setcopyright{acmcopyright}
\copyrightyear{2021}
\acmYear{2021}

\acmConference[ESEM 2021 Registered Reports]
  {15th ACM/IEEE International Symposium on Empirical Software Engineering and Measurement}
  {October 11--15, 2021}
  {Bari, Italy}
\acmBooktitle{ESEM 2021: 15th ACM/IEEE International Symposium on Empirical Software Engineering and Measurement, October 11--15, 2021, Bari, Italy}
\acmPrice{15.00}
\acmISBN{978-1-4503-XXXX-X/21/10}

\begin{document}

\title[\SHORTTITLE]{\TITLE}

\author{Amador Durán}
\affiliation{%
    \institution{\SCORE, \ITRESUS}
    \institution{\US}
    \city{\Sevilla}
    \country{\Spain}
}
\email{amador@us.es}

\author{Pablo Fernández}
\affiliation{%
    \institution{\SCORE, \ITRESUS}
    \institution{\US}
    \city{\Sevilla}
    \country{\Spain}
}
\email{pablofm@us.es}

\author{Beatriz Bernárdez}
\affiliation{%
    \institution{\ITRESUS}
    \institution{\US}
    \city{\Sevilla}
    \country{\Spain}
}
\email{beat@us.es}

\author{Nathaniel Weinman}
\affiliation{%
		\institution{\ACE}
    \institution{\UCB}
    \city{\Berkeley}
    \country{\USA}
}
\email{nweinman@berkeley.edu}

\author{Asl\i{} Akal\i{}n}
\affiliation{%
		\institution{\ACE}
    \institution{\UCB}
    \city{\Berkeley}
    \country{\USA}
}
\email{asliakalin@berkeley.edu}

\author{Armando Fox}
\affiliation{%
		\institution{\ACE}
    \institution{\UCB}
    \city{\Berkeley}
    \country{\USA}
}
\email{fox@berkeley.edu}
%

\renewcommand{\shortauthors}{Durán \etal}




\begin{abstract}
\emph{Context}. 
\PP has been found to increase student interest in \CS, particularly so for women, and would therefore appear to be a way to help remedy the under--represen\-ta\-tion of women in the field. However, one reason for this under--represen\-ta\-tion is the unwelcoming climate created by gender stereotypes applied to engineers in general, and to software engineers in particular, assuming that \change{men perform better than their women peers.} If this same bias is present in \pp, it could work against the goal of improving gender balance in computing. %
\noindent\emph{Objective.} 
\change{In a remote setting in which students cannot directly observe the gender of their peers,} we aim to explore whether \SE students behave differently when the \emph{perceived} gender of their remote \pp partners changes, searching for differences in (i) the perceived productivity compared to solo programming; (ii) the partner's perceived technical competency compared to their own; (iii) the partner's perceived skill level; (iv) the interaction behavior, such as the frequency of source code additions, deletions, validations, etc.; and (v) the type and relative frequencies of dialog messages used for collaborative behavior in a chat window. Although there are some studies on \pp performance and gender pair combination, to the best of our knowledge there are no studies on the impact of gender stereotypes and bias \emph{within} the pairs themselves. %
\noindent\emph{Method}. 
We have developed an online platform (\twincode) that randomly classifies students into gender--balanced groups, arranges them in pairs for remote \pp (sharing an editor window and a chat window), and can selectively deceive one or both partners regarding the gender of the other via the use of a clearly gendered avatar. %
Several behaviors are automatically measured during the \pp process, together with two questionnaires and a semantic tagging of the pairs' conversations. %
We will perform a series of 
experiments to identify the effect, if any, of possible gender bias in remote \pp interactions. %
Students in the control group will have no information about their partner's gender; students in the treatment group will receive such information but will be selectively deceived about their partner's true gender.
To analyze the data, apart from checking reliability of questionnaire data using Cronbach's alpha and Kaiser criterion, for each response variable we will %
(i) compare control and experimental groups for the score distance between two in--pair tasks; then, using the data from the experimental group only, we will %
(ii) compare scores using the partner's perceived gender as a within--subjects variable; and 
(iii) analyze the interaction between the partner's perceived gender (within--subjects) and the subject's gender (between--subjects). %
For the (i) and (ii) analyses we will use t--tests, whereas for the (iii) analyses we will use mixed--model \ANOVA{}s. 

\end{abstract}


\maketitle



\section{Introduction} \label{sec:introduction}

\PP is an increasingly popular collaboration paradigm that has been shown to be an effective tool in \CS education as measured by positive influence on grades, class performance, confidence, productivity, and motivation to stay 
\cite{dpp-survey-2015}, especially for women \cite{Rodriguez2017, werner-et-al}. %
In \pp, two partners work closely together to solve a programming task. As such, their ability to engage with each other is key. %
However, these interactions are influenced by implicit gender bias \cite{Hofer2015,Martell1996}, such as assuming 
\change{women are} less technically competent \cite{Martell1996}. %
This is a widely observed phenomenon even in highly--structured settings \cite{jarratt-iticse-2019,dpp-survey-2015}. %
Social sciences research indicates that one’s behavior of an individual is affected by the behavior of their peers \cite{Eckles7316}. %
Therefore, implicit gender bias based on perception of peers may have effects on one’s behavior, potentially influencing \pp experience.

In this work, \change{in a non--colocated (i.e.\ remote) environment in which the gender of the peers cannot be directly observed,} our goal is to explore whether \SE students change their
behavior when the \emph{perceived} gender of their remote \pp partners changes \change{from man to woman or vice versa}. %
%
\change{Note that, while we recognize that many 
students may identify as neither men nor women, our initial exploration focuses primarily on interactions between students who identify as one of these, so that we can better align our findings with the existing literature on implicit gender bias. The potential biases in interactions involving gender-fluid, non-gender-conforming, or nonbinary students is a rich and complex topic deserving its own subsequent study.}



\begin{figure*}[b!]
	\centering
  \begin{tabular}{>{\hspace{-0.5em}}c>{\hspace{1.25em}}c}
    \includegraphics[width=\columnwidth]{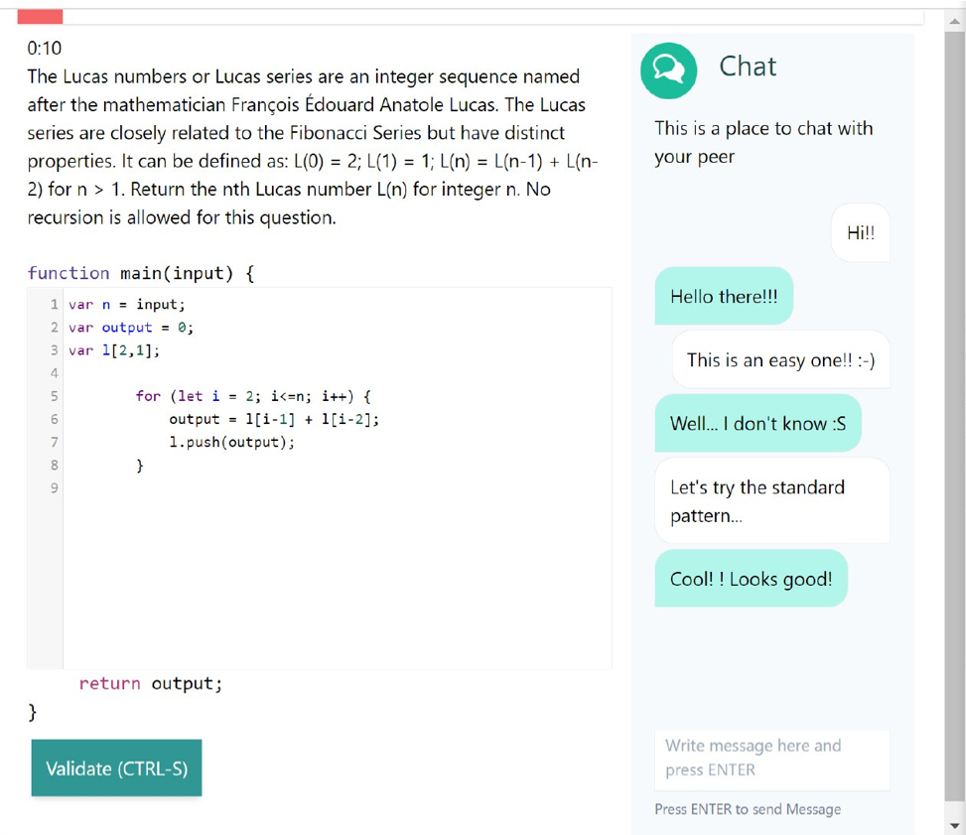} &
    \includegraphics[width=\columnwidth]{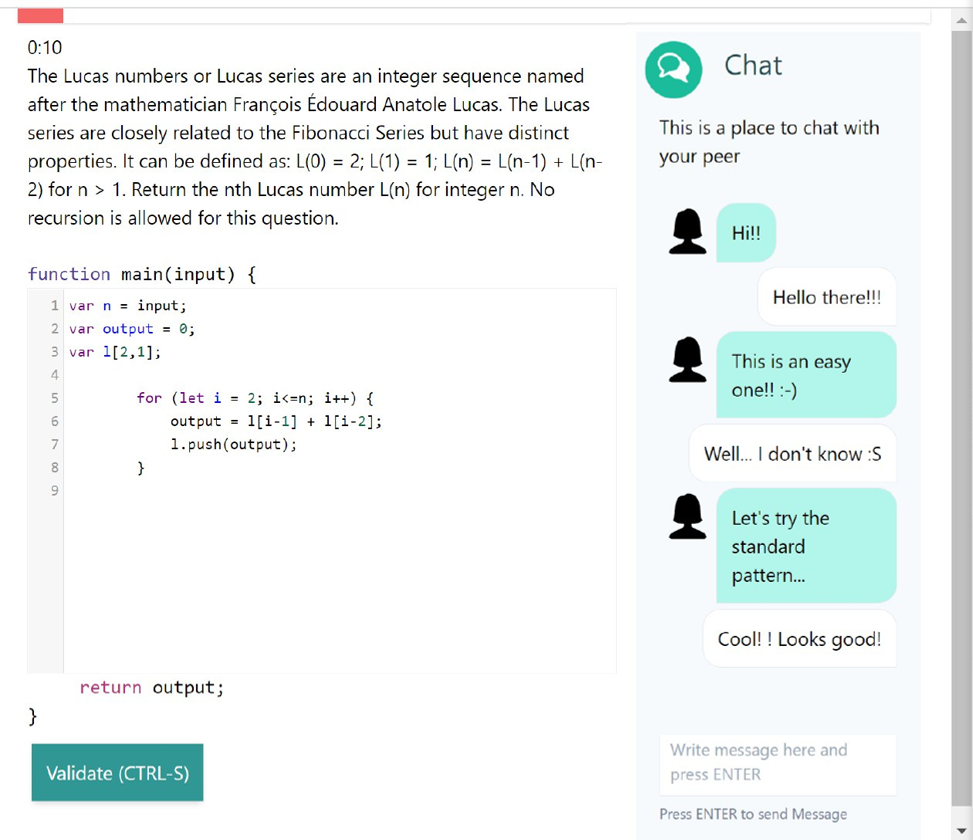}  
  \end{tabular}
	\caption{\twincode user interface for control group (left) and experimental group (right)}
	\label{fig:twincode}
\end{figure*}

To achieve our goal, we plan to search for differences not only in the perceived productivity of \pp compared to solo programming, the partner's perceived technical competency compared to their own, and the partners' perceived skill level, but also in the interaction behavior, i.e. the frequency of source code additions, deletions, validations, etc., and the type and relative frequency of dialog messages used for collaborative behavior. %

To get early feedback on the infrastructure supporting our proposal, we ran two pilot studies, one at each university, 
with a limited number of students, where we could check the comprehensibility of the questionnaires used to gather subjective data, the applicability of the message tagging (described in Section \ref{sec:research_questions}), and the capabilities of the \twincode platform, which is briefly described below. 

%


\subsection{The \twincode platform} \label{sec:twincode}


To support our study, we have developed the \twincode remote \pp platform, which %
manages the registration of students, the random allocation to gender-balanced groups, the random allocation into pairs, the random assignment of programming exercises to pairs, and the automatic collection of interaction metrics and dialog messages. %

As shown in \figurename~\ref{fig:twincode}, \twincode offers a source code editor where the students concurrently develop the solution to a proposed exercise and can validate it against several test cases. %
It also offers a chat window, where they can collaborate to solve the exercise. Note that a gendered avatar is displayed for the student in the experimental group only (right), but not for the one in the control group (left). %



\subsection{Related Work} \label{sec:related_work}



\begin{table*}
	\centering
	\small
  \renewcommand{\arraystretch}{1.25}
	\begin{tabularx}{\textwidth}{l>{\hsize=.25\hsize}X>{\hsize=.66\hsize}XX}
		\hline\noalign{\smallskip}

		\thead{Reference}       & 
		\thead{Object of study} & 
		\thead{Metrics}         & 
		\thead{Findings}        \\		

    \noalign{\smallskip}\hline\noalign{\smallskip} 
		
    Katira \etal \cite{Katira-icse-2005} & 
    Compatibility of student pair programmers & 
    Web--based peer evaluation survey that required the students to evaluate the contributions of their partner and the pair compatibility as perceived by the student & 
    Students are compatible with partners whom they perceive of similar skill. Mixed-gender pairs are less likely to report compatibility. \\
	
    Choi \etal \cite{choi-bit-2015} & 
    PP gender combinations & 
    Productivity, quality of source code, compatibility and communication between pairs &

    Significant differences in the levels of pair compatibility and communication between the same gender pair type, \change{woman--woman} and \change{man--man}.\\

    Gómez \etal \cite{Gomez2017} & 
    PP gender combinations & 
    Productivity & 
    Similar productivity rates for the three gender pair combinations. The programming assignments had a significant impact on the productivity. Greater variability of productivity rates with mixed gender pairs (\change{man--woman}) 
    was observed. \\	
   
    Jarrat \etal \cite{jarratt-iticse-2019} & 
    PP gender combinations & 
    Weekly attendance, work accomplished during lab and perceptions of productivity & 
    Students who were randomly assigned a \change{woman as a} 
    partner (rather than a \change{man as a} 
    partner) attended class more often, were more confident that the solution was correct, and more confident in the finished product that they created. However, being assigned a \change{woman as a} 
    partner was also associated with completing a smaller percentage of the assignment. \\

    \noalign{\smallskip}\hline 
    
    \end{tabularx}
	
	\caption{Empirical studies about gender in pair programming}
	\label{tab:related_work}
\end{table*}

Several literature reviews \cite{salleh-2010-effects,hanks-2011-pair,Gupta-2021} have compiled the empirical research on using \pp in higher education, \change{with \cite{dpp-survey-2015} being focused on distributed pair programming from a teaching perspective}. \change{By means of controlled experiments, %
remote and co--located \pp are compared in \cite{stotts-2003-virtual,al-2016-effectiveness}, showing comparable results.} %
In most of the cases, the analyzed variables are related to performance in terms of time, quality, or code tests passed. Students perceptions have been also analyzed in terms of confidence, satisfaction, motivation, or personality \cite{salleh-2014-investigating}.

\tablename~\ref{tab:related_work} summarizes the empirical studies on the influence of gender in \pp, including findings such as (i) same--gender pairs are more ``democratic''; (ii) women working in pairs were more confident than those working solo; and (iii) in mixed-gender, pairing women particularly do not benefit \cite{Gupta-2021}. Although such studies reveal that gender seems to be a key factor in \pp, none of them study gender bias in \pp.

Many factors other than gender may affect the outcomes of remote programming sessions \cite{Chaparro2005FactorsAT,Thomas2003}. Previous research on productive pairing looked at factors such as skill levels, autonomy in choosing one’s partner \cite{Xinogalos2017StudentPO}, and different personalities \cite{personality-hannay2010}. %
Nevertheless, the work on gender composition of pairs found conflicting results about whether same-gender or mixed-gender pairings are more effective \cite{choi-bit-2015,genderinpp-choi2013,Hofer2015,kuttal2019}. One possible explanation is that gender correlates with other dimensions that may affect the pairs' collaboration, but these correlations may vary between different environments. For example, women in a class may, on average, have higher skill level than men because they had to face more societal barriers to enter the class. On the other hand, they may, on average, have lower skill level if women with no background are more actively recruited.

\section{Research Questions} \label{sec:research_questions}

\change{Our study is based on the hypothesis that gender bias will
  lead to observable differences based on subjects' perceptions of the
  gender of their partners, i.e., they will score men and women
  differently for similar tasks and also behave differently depending
  on the perceived gender of their partner.} %
%
%
To study our hypothesis, we plan to apply methodological triangulation \cite{Denzin}, using several methods to collect data and approaching a complex phenomenon like human behavior from more than one standpoint \cite{Cohen}. %
%
In our case, three different data sources will be used: questionnaires completed by the subjects, data collected automatically by the \twincode platform, and data produced by \change{two different experimenters} analyzing and tagging dialog messages \change{and cheking interrater agreement using Cohen's kappa coefficient \cite{cohens-kappa}}.

With respect to the data collected using questionnaires, our research questions are:

\begin{enumerate}[label=\textbf{RQ$_{\arabic*}$}]

\item 
In remote \pp, does gender bias affect perceived productivity compared to solo programming?

\item
In remote \pp, does gender bias affect the partner's perceived technical competency compared to one's own technical competency?

\item
In remote \pp, does gender bias affect how partners' skills are perceived?

\end{enumerate}

With respect to the data automatically collected by the \twincode platform---which could be increased in the future---our research question is:

\begin{enumerate}[resume,label=\textbf{RQ$_{\arabic*}$}]

\item 
In remote \pp, does gender bias affect the frequencies or relative frequencies with which each partner produces source code additions, source code deletions, successful validations, failed validations, and dialog messages?

\end{enumerate}



\begin{table}[t]
  \centering
  \small
  \renewcommand{\arraystretch}{1.25}  
  \begin{tabularx}{\columnwidth}{cXX}
    \hline\noalign{\smallskip}
    
    \thead{Tag} & \thead{Description} & \thead{Examples} \\
    
    \noalign{\smallskip}\hline\noalign{\smallskip}   
    
    I &
    Informal &
    \emph{LOL! Hahaha!} \\
        
    F &
    Formal &
    All messages except informal \\

		\hline\noalign{\smallskip}        

    S & 
    Statement of information or explanation &
    \emph{We need to create a program for kids to learn math} \\
    
    U & 
    Opinion or indication of uncertainty &
    \emph{Unsure how to add strings together} \\
    
    D & 
    Explicit instruction &
    \emph{Wait put the if back} \\
    
    SU & 
    Polite or indirect instruction &
    \emph{Maybe we can do if user choice = +} \\
    
    ACK & 
    Acknowledgement &
    \emph{Oh ok gotcha} \\
    
    M & 
    Meta--comment or reflection &
    \emph{Hmmm} \\
    
    QYN & 
    Yes/no question &
    \emph{Can the answer be negative?} \\
    
    QWH & 
    Wh-- question (who, what, where, when, why, and how) &
    \emph{How do I take in their input?} \\
    
    AYN & 
    Answer to yes/no question &
    \emph{Yea} \\
    
    AWH & 
    Answer to wh-- question &
    \emph{The program should be able to generate erroneous questions} \\
    
    FP & 
    Positive task feedback &
    \emph{Oh nice} \\
    
    FNON & 
    Non--positive task feedback &
    \emph{Thats weird} \\

    O & 
    Off--task &
    \emph{Wow its sweet in this room} \\

		\hline\noalign{\smallskip}
  \end{tabularx}
  \caption{Dialogue tags from \cite{Rodriguez2017} augmented with orthogonal informal/formal tags
  }
  \label{tab:tags}
\end{table}

The manual semantic tagging of the dialog messages classifies each message into two orthogonal dimensions. The first dimension uses the 13 tags proposed in \cite{Rodriguez2017} (tags from \level{S} to \level{O} in \tablename~\ref{tab:tags}). The second dimension classifies each message as \level{formal} or \level{informal}. %
With respect to this data source, our research questions are:

\begin{enumerate}[resume,label=\textbf{RQ$_{\arabic*}$}]

\item 
In remote \pp, does gender bias affect the frequency or relative frequency of the different types of dialog messages?

\item 
In remote \pp, does gender bias affect the relative frequency of formal and informal dialog messages?

\end{enumerate}



\section{Variables} \label{sec:variables}

In this section, we describe all the variables we will consider in our study. Note that depending on the development of the \twincode platform, more automatically measured dependent variables could be added in the future.

When used, abbreviations are enclosed in parentheses after variables' names.

\subsection{Independent Variables} \label{sec:independent_variables}

\begin{description}[style=unboxed]

\item[\variable{group}]
nominal factor representing the group (\level{experimental} or \level{control}) subjects are randomly allocated to.

\item[\variable{time}]
nominal factor representing the moment (\level{t$_1$} and \level{t$_2$}) in which the first and second in--pair tasks are performed by the subjects.

\item[\change{induced partner's gender (\variable{ipgender})}]
\change{nominal factor representing the induced partner's \change{binary} gender
  (\level{man} or \level{woman} for the experimental group, and
  \level{none} for the control group, \change{in which gender is not revealed}) during the in--pair tasks. This variable, which is directly related to the gender bias treatment, is operationalized by means of the instructions provided at the beginning of the tasks (``\ldots work with your partner. She is \ldots'') and by the gendered avatar that is visible during the in--pair tasks for the experimental group and that is swapped between tasks. Subjects in the control group receive no treatment, i.e., they do not see any information about the gender of their partners in any way, neither textual nor graphical.}

\item[\variable{gender}]
nominal factor representing subject's gender, \change{which may be
  \level{man}, \level{woman}, or any other option as freely expressed in the demographic form during registration.}

\end{description}

\subsection{Dependent Variables} \label{sec:dependent_variables}

The response variables measured using questionnaires containing 
0--10 linear numerical response items are the following:

\begin{description}[style=unboxed]

\item[Perceived productivity compared to solo programming (\variable{pp})]
interval variable measuring the subject's perceived productivity compared to solo programming after each in--pair task (see \RQ{1}). Low values correspond to better solo programming productivity, i.e., ``solo programming would have been more productive than pair programming'', whereas high values correspond to better pair programming productivity, i.e. ``pair programming has been more productive than solo programming''.

\item[Perceived partner's technical competency compared to their own (\variable{pptc})]
interval variable measuring the subject's partner's perceived technical competency compared to their own after each in--pair task (see \RQ{2}). Low values correspond to higher subject's productivity, i.e., ``I have been more productive than my partner'', whereas higher values correspond to high partner's productivity, i.e. ``My partner has been more productive than me''.

\item[Compared partners' skills (\variable{cps})]
interval variable measuring whether the subject perceived better skills in \change{their} first or second partner in the in--pair tasks (see \RQ{3}). Low values correspond to the first partner, i.e., ``My first partner was a better partner than my second partner'', whereas high values correspond to the second partner, i.e. ``My second partner was a better partner than my first partner''. %
\change{In the case of the experimental group only,} this variable is
transformed after collection using an R script \change{in such a way
  that} low values correspond to the partner perceived as a
\change{man}, and high values to the partner perceived as a \change{woman, in order to analyze whether there is a gender bias in the scoring.}

\end{description}

\change{Apart from the variables described above, the questionnaires will also include questions about the perceived gender of their parners at each task. The corresponding variable is described below:}

\begin{description}[style=unboxed]

\item[\change{perceived partner's gender (\variable{ppgender})}]
\change{nominal factor representing the \change{subject's perception
    of their partner's gender} (\level{woman}, \level{man}, \level{I don't know}, or \level{I don't remember}) at each in--pair task.}

\end{description}  

The response variables automatically collected by the \twincode platform and related to the interaction during the in--pair programming exercises (see \RQ{4}) are listed below. %
Every variable \variable{v} represents a frequency, i.e., a count, and its associated relative frequency is computed with respect to the the sum of the frequencies of the two subjects in a pair. %
For example, let us suppose that subjects $i$ and $j$ are the two members of a pair, and \variable{v$_i$} and \variable{v$_j$} are the corresponding values of the \variable{v} variable. %
In this case, the relative frequencies for each subject would be $\variable{v}_{i} \over{\variable{v}_{i} + \variable{v}_{j}}$ and $\variable{v}_{j} \over {\variable{v}_{i} + \variable{v}_{j}}$, respectively. 

\begin{description}[style=unboxed]

\item[source code additions (\variable{sca})]
Ratio scale variable representing the count of characters added by a subject to the source code window during an in--pair task.

\item[source code deletions (\variable{scd})]
Ratio scale variable representing the count of characters deleted by a subject from the source code window during an in--pair task.

\item[successful validations (\variable{okv})]
Ratio scale variable representing the count of successful validations of the source code performed by a subject during an in--pair task.

\item[unsuccessful validations (\variable{kov})]
Ratio scale variable representing the count of unsuccessful validations of the source code performed by a subject during an in--pair task.

\item[dialog messages (\variable{dm})]
Ratio scale variable representing the count of dialog messages sent by a subject during an in--pair task.

\end{description}

The response variables related to the manual tagging of the dialog messages (see \RQ{5} and \RQ{6}) correspond to the  tags in \tablename~\ref{tab:tags} and are listed below. %
%
Every variable represents a frequency, i.e., a count, and its associated relative frequency is computed with respect to the number of dialog messages generated by the subject during an in--pair task, which is defined by the \variable{dm} variable specified above. %

\newcommand{\tagvar}[1]{Ratio scale variable representing the count of #1 messages generated by a subject during an in--pair task.}

\begin{description}[style=unboxed]

\item[\variable{i}]
\tagvar{informal}

\item[\variable{f}]
\tagvar{non--informal, i.e. formal,}

\item[\variable{s}]
\tagvar{statement of information or explanation}

\item[\variable{u}]
\tagvar{opinion or indication of uncertainty}

\item[\variable{d}]
\tagvar{explicit instruction}

\item[\variable{su}]
\tagvar{polite or indirect instruction}

\item[\variable{ack}]
\tagvar{acknowledgment}

\item[\variable{m}]
\tagvar{meta--comment or reflection}

\item[\variable{qyn}]
\tagvar{yes/no question}

\item[\variable{qwh}]
\tagvar{wh- question (who, what, where, when, why, and how)}

\item[\variable{ayn}]
\tagvar{answer to yes/no question}

\item[\variable{awh}]
\tagvar{answer to wh- question}

\item[\variable{fp}]
\tagvar{positive task feedback}

\item[\variable{fnon}]
\tagvar{non--positive task feedback}

\item[\variable{o}]
\tagvar{off--task}

\end{description}


\subsection{Confounding Variables}

The confounding variables that will be controlled during the experiment are the following.

\begin{description}

\item[Subject's technical skills]
To control the variability caused by each subject on \change{their} partner, pairs are kept the same during the entire experiment, although the subjects are not informed about this fact until the end of the experiment. %
Ideally, this would make the conditions of the two in--pair tasks the same except for the programming exercises (see below) and for the perceived gender in the case of the experimental group.

\item[Programming exercises]
In order to avoid potential differences among the programming exercises used during in--pair tasks, they \change{are of similar difficulty and} are randomly assigned.


\end{description}



\section{Participants} \label{sec:participants}

As we have done in the pilot studies, our plan is offering our students the possibility to participate in the \twincode study. %
At the University of Seville, the participants will be 3$^\text{rd}$--year students of the Degree in Software Engineering. We expect the number of participants to be around 150--200. %

At the University of California Berkeley, the participants will be students enrolled in the 1$^\text{st}$ and 2$^\text{nd}$ semester \CS courses. In this case, we expect the number of participants to be around 100--300.



\section{Execution Plan} \label{sec:execution}

We plan to perform a baseline experiment at the University of Seville in the first semester of the 2021--2022 academic course, and then, replicate the experiment at the University of California Berkeley at the beginning of the second semester. \change{At each location, the experimental material provided to the students will be localized, i.e., in Spanish in Spain and in English at the USA, and the translation will be carefully checked by bilingual experimenters.}\footnote{At the time of writing, discussions are underway for another replication at the Northern Technical University at Ecuador. \change{If this replication is eventually carried out, the experimental material will also be adapted to the local variant of Spanish if deemed necessary.}}

Independently of being the baseline or the replication, the execution plan of the \twincode study consists of the steps described below.

\subsection{Recruitment}

In this initial step, we plan to motivate the students to voluntarily participate in the study as an interesting experience in remote pair programming but without mentioning that the main goal is to study the effect of gender bias. %
We also remark that for the purpose of the study, they must remain anonymous to their partners, so they must neither mention nor ask any personal information.

The interested students must register in the \twincode platform providing some demographic data and accepting the participation conditions.

\subsection{Training}

One week or so before the experiment execution, we plan to provide a short training session about the \twincode platform, so students become familiar with it and any concerns about how to use the platform were dispelled, thus reducing any potential anxiety for using a new platform.



\begin{figure}
	\centering
		\includegraphics[width=0.95\columnwidth]{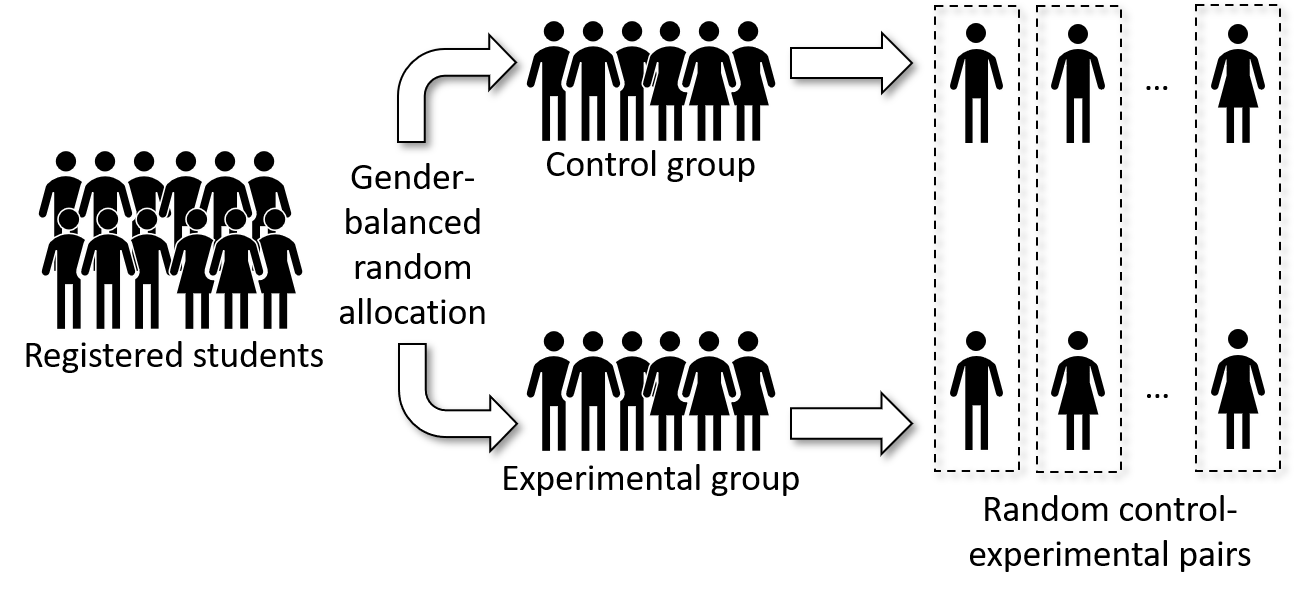}
	\caption{Experimental process (subject allocation to groups)}
	\label{fig:allocation}
\end{figure}

\subsection{Experiment Execution}

For experiment execution, all registered students must log into the \twincode platform, which will automatically allocate them into the control and experimental groups. This allocation is random and gender--balanced, \change{i.e., the proportion of men and women in each group is as close as possible}. %
Once all students are allocated to groups, they are randomly allocated into control--experimental pairs, as shown in \figurename~\ref{fig:allocation}. 

\begin{figure}
	\centering
		\includegraphics[width=\columnwidth]{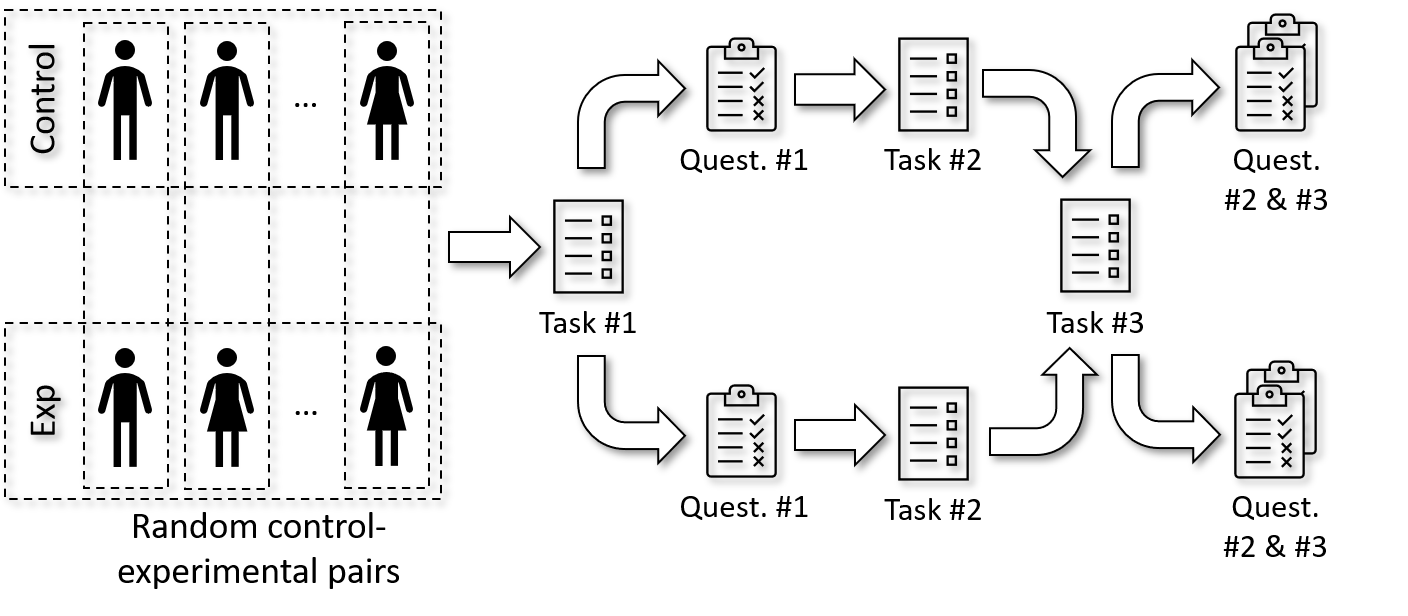}
	\caption{Experimental process (tasks)}
	\label{fig:tasks}
\end{figure}

After subject allocation, the pairs are presented a programming exercise  that they must solve collaboratively (labeled as \variable{Task \#1} in \figurename~\ref{fig:tasks}) in the \twincode platform. %
They are given 20 minutes to do it, and in case they finished earlier, another exercise of increasing difficulty is presented. At the end of the 20--minute period, they are asked to individually fill in a questionnaire about the perceived productivity compared to solo programming and about the perceived partner's technical competency compared to their own. They are given 10 minutes to fill in the questionnaire.

After filling the first questionnaire, the students are presented another programming exercise to be solved individually in 10 minutes (\variable{Task \#2} in \figurename~\ref{fig:tasks}). As in the previous task, another exercise of increasing difficulty is presented if they finish earlier. %
The main purpose of this individual task is to make students forget about their first partners, i.e. their way of writing dialog messages or source code, so they do not recognize them in the second in--pair task.

After the individual task, pairs are presented again a new
collaborative programming exercise that they must solve in similar
conditions to the exercise in \variable{Task \#1}. In this second
in--pair exercise, the avatar gender \change{(which in this initial
  study will always indicate binary gender)} is swapped with respect to the first exercise for the subjects in the experimental group. %
%
\change{Note that pairs are kept the same in order to reduce the variability due to the subjects themselves, which could possibly have had a confounding effect in case of a new pair allocation for \variable{Task \#3}.}

Once Task \#3 is finished, students are asked to fill in the same questionnaire than they filled after Task \#1, and another questionnaire comparing the \change{perceived genders} and skills of the first and second partners. They are given 15 minutes for responding both questionnaires.

Finally, they are informed about the actual goal of the study and that the pairs remained the same during the experiment. At this moment, they are allowed to withdraw their data if they wish to do so. 

\subsection{Data Analysis}



\change{During the manual tagging of the dialog messages, all pairs in which the gender of any of the peers is disclosed in any way are excluded from the data analysis.} %
Then, before analyzing response variables, the internal consistency of the questionnaire data is checked using Cronbach's alpha and Kaiser criterion. %
After that, for every dependent variable \variable{v}, we compute the distance between the two in--pair tasks as the absolute value of the difference, i.e. |\,\variable{v}(t$_2$) -- \variable{v}(t$_1$)\,|. Ideally, this distance should be lower for the students in the control group (no information about partners' genders) than for those in the experimental group (with two different perceived partners' genders at \level{t$_1$} and \level{t$_2$}). %
Therefore, for every variable, we perform a t--test to detect distance differences between the groups.

Then, using the data from the experimental group only, we perform a t--test 
to detect differences in the scores of every dependent variable
between perceived partner's gender, i.e. to detect differences in the
scores when partners are perceived as \change{men vs.\ as women}. %
Finally, to detect a potential interaction between the perceived partner's gender and the subject's gender, we perform a mixed--model \ANOVA with the perceived gender as a within--subjects variable and subject's gender as a between--subjects variable. 
%
\change{As complementary analyses, we also study (i) the correlation between the induced and the perceived gender for the subjects in the experimental group, and the distribution of the perceived gender (if any) in the control group; and (ii) the potential cultural impact of the different locations at which the experiment is carried out.}
 
All the data analysis will be performed using R scripts, that will be available in a public repository together with the datasets in the corresponding laboratory package.

\begin{acks}
We would like to thank the students who volunteered to participate in the pilot studies at the University of California, Berkeley and the University of Seville. %
We particularly acknowledge Vron Vance's assistance regarding inclusive language around gender identity.

\FUNDING
\end{acks}


\bibliographystyle{ACM-Reference-Format}
\bibliography{duran_esem_rr_2021}


\end{document}